\documentclass{article}

\usepackage{amsmath,amssymb}
\usepackage{graphicx}

\begin{document}
\baselineskip=20pt
\hsize=340pt
\vsize=490pt
\begin{titlepage}
\title{
\hfill\parbox{4cm}
{\normalsize RIKEN-TH-170
}\\
\vspace{1em}
{\bf 
Numerical Evaluation of Gauge Invariants for $a$-gauge Solutions\\ 
in Open String Field Theory
}
\vspace{5ex}
}

\author{
Isao {\sc Kishimoto}$^{1}$\footnote{E-mail address:
ikishimo@riken.jp}
\ and
Tomohiko {\sc Takahashi}$^{2}$\footnote{E-mail address:
tomo@asuka.phys.nara-wu.ac.jp}
\\
\vspace{2ex}\\
{\it
$^1$Theoretical Physics Laboratory, RIKEN,}\\
{\it 
Wako 351-0198, Japan}\\
{\it
$^2$Department of Physics, Nara Women's University,}\\
{\it 
Nara 630-8506, Japan}\\
}

\date{\normalsize October, 2009}
\maketitle
\thispagestyle{empty}

\begin{abstract}
\normalsize

We evaluate gauge invariants, action and gauge invariant overlap,
for numerical solutions which satisfy the ``$a$-gauge'' condition
with various values of $a$ in cubic open bosonic string field theory.
We use the level truncation approximation and an iterative procedure to
 construct numerical solutions in the twist even universal space.
The resulting gauge invariants are numerically stable and almost equal
 to those of Schnabl's solution for tachyon condensation.
Our result provides further evidence that these numerical and analytical
 solutions are gauge equivalent. 

\end{abstract}

\end{titlepage}

\section{Introduction}

The string field theory is expected to be a candidate for nonperturbative
formulation of the string theory. 
The study of solutions of string field theories is important
for understanding nonperturbative phenomena in the string theory.
In particular, since Schnabl constructed an
analytic solution \cite{Schnabl:2005gv} for tachyon condensation
 in the framework of
cubic open bosonic string field theory, there have been a
number of new developments in this field.

A prominent feature of Schnabl's solution 
is the potential height at the solution,
which is given by the evaluation of the action and
is proved to be equal to the tension of D25-brane analytically
as is consistent with Sen's conjecture.
Other gauge invariant observable, which is called gauge invariant
overlap,  was calculated for Schnabl's solution with an analytic
method \cite{Ellwood:2008jh, Kawano:2008ry} and with the $L_0$-level
truncation approximation \cite{Kawano:2008ry}. The value suggests that
 the solution may be related to the boundary state for 
D-brane \cite{Ellwood:2008jh, Kawano:2008jv}.

On the other hand, before advent of Schnabl's analytic solution, the
numerical solution in the Siegel gauge was constructed using level
truncation approximation and its potential height was evaluated
\cite{Sen:1999nx, Moeller:2000xv, Gaiotto:2002wy}.
It is almost the same as D-brane tension.
In \cite{Kawano:2008ry}, the gauge invariant overlap for the numerical
solution was computed and it turned out that the value is almost the
same as that of the analytic solution.
These results are consistent with the expectation that these solutions
may be gauge equivalent. 

Actually, there are other numerical solutions.
Here, we focus on the solutions in Asano-Kato's $a$-gauge
\cite{Asano:2006hk} which was proposed as a consistent gauge fixing
condition with a real parameter $a$ corresponding to the covariant gauge
in the conventional gauge theory.
Using this gauge, numerical solutions for tachyon condensation were
constructed and their potential heights were evaluated
 with level truncation up to level $(6,18)$ in \cite{Asano:2006hm}.
We construct numerical solutions in the $a$-gauge for various $a$ 
for higher level and evaluate gauge invariants, action and gauge
invariant overlap, for them \cite{Kishimoto:2009cz}.

It turned out that the values for each configuration approach 
those of the analytic solution with increasing level.
This fact implies that these numerical solutions
in the various $a$-gauges, not only in the Siegel gauge, 
are all gauge equivalent to the Schnabl's analytic one.
Namely, these various solutions may represent a unique nonperturbative
tachyon vacuum in bosonic string field theory.
Furthermore, our numerical results may indicate that the
level truncation approximation 
not only in the Siegel gauge but also in the $a$-gauge
would be reliable in order to investigate nonperturbative
vacuum in bosonic string field theory.

The rest of this article is organized as follows. First, we review
gauge invariant overlap in \S \ref{sec:GIO} and $a$-gauge condition in
\S \ref{sec:AKa-gauge}. Then we explain our method to construct
numerical solutions in \S \ref{sec:method}. In \S \ref{sec:results}, 
we display our numerical results. 
Finally, we give conclusion in \S \ref{sec:summary}.

\section{Gauge invariant overlap
\label{sec:GIO}}

The gauge invariant overlap ${\cal O}_V(\Psi)$ is defined by contraction
of an open string field $\Psi$ and an on-shell closed string state.
More precisely, ${\cal O}_V(\Psi)$ can be expressed as\footnote{
See, for example, \cite{Kawano:2008ry} for details.}
\begin{eqnarray}
 {\cal O}_V(\Psi)&=&\langle \hat\gamma (1_{\rm
  c},2)|\Phi_V\rangle_{1_{\rm c}}|\Psi\rangle_2,
\label{eq:Q_VST}
\end{eqnarray}
where $\langle \hat\gamma (1_{\rm c},2)|$ is the Shapiro-Thorn vertex
\cite{Shapiro:1987ac} and $|\Phi_V\rangle$ is 
given by a  matter primary field $V_{\rm m}(z,\bar z)$
with dimension $(1,1)$:
$|\Phi_V\rangle= c_1\bar c_1V_{\rm m}(0,0)|0\rangle$.
Using the relations,
${\cal O}_V(Q_{\rm B}\Lambda)=0$ and 
${\cal O}_V(\Psi *\Lambda)={\cal O}_V(\Lambda*\Psi)$,
one can see that ${\cal O}_V(\Psi)$ is gauge invariant:
$\delta_{\Lambda}{\cal O}_V(\Psi)=0$
under the gauge transformation of string field,
$\delta_{\Lambda}\Psi = Q_{\rm B}\Lambda + \Psi *\Lambda
- \Lambda*\Psi$,
which leaves the action
\begin{eqnarray}
 S(\Psi)&=&-\frac{1}{g^2}\left(
\frac{1}{2}\langle \Psi,Q_{\rm B}\Psi\rangle+\frac{1}{3}\langle
\Psi,\Psi*\Psi\rangle\right)
\label{eq:action}
\end{eqnarray}
invariant.

Let us evaluate the gauge invariant overlap for Schnabl's solution for
tachyon condensation $\Psi_{\rm Sch}$ \cite{Schnabl:2005gv}, which can
be expressed as
\begin{eqnarray}
\label{eq:Schnabl_sol}
 \Psi_{\rm
  Sch}
&=&\psi_0+\sum_{n=0}^{\infty}(\psi_{n+1}-\psi_n-\partial\psi_r|_{r=n})
\end{eqnarray}
with a particular string field $\psi_r$:
\begin{eqnarray}
 \psi_r=\frac{2}{\pi}U^{\dagger}_{r+2}U_{r+2}
\left[\frac{-1}{\pi}({\cal B}_0+{\cal B}_0^{\dagger})
\tilde c(\tilde x_r)
\tilde c(-\tilde x_r)
+\frac{1}{2}(\tilde c(\tilde x_r)+\tilde c(-\tilde x_r))
\right]|0\rangle,
\end{eqnarray}
where $\tilde x_r=\pi r/4$, $U_r=(2/r)^{{\cal L}_0}$,
${\cal B}_0=b_0+\sum_{k=1}^{\infty}\frac{2(-1)^{k+1}}{4k^2-1}b_{2k}$,
${\cal L}_0=\{Q_{\rm B},{\cal B}_0\}$ and $\tilde c(\tilde z)=(\cos
\tilde z)^2 c(\tan\tilde z)$.
Using the fact that ${\cal O}_V(\psi_r)$ is independent of $r$, we
have \cite{Ellwood:2008jh,Kawano:2008ry, Kawano:2008jv}
\begin{eqnarray}
{\cal O}_V(\Psi_{\rm Sch})&=&{\cal O}_V(\psi_0)
=\frac{1}{2\pi}\langle B|c_0^-|\Phi_V\rangle,
\label{eq:O_VPsi=BPhi}
\end{eqnarray}
where $\langle B|$ is the boundary state for D-brane.
In order to get nonzero value for zero momentum open string fields,
we take the dilaton state with zero momentum as an
on-shell state: $\Phi_V=-\frac{1}{26}\eta_{\mu\nu}\alpha_{-1}^{\mu}
\bar\alpha_{-1}^{\nu}c_1\bar c_1|0\rangle$.
Then (\ref{eq:O_VPsi=BPhi}) gives
\begin{eqnarray}
 {\cal O}_V(\Psi_{\rm Sch})/V_{26}=\frac{1}{2\pi},
\label{eq:O_eta=1/2pi}
\end{eqnarray}
where $V_{26}$ is the volume factor.

\section{Asano-Kato's $a$-gauge
\label{sec:AKa-gauge}}

The $a$-gauge condition, proposed by Asano and Kato in \cite{Asano:2006hk},
in the classical sector, namely the worldsheet ghost number one
sector,\footnote{
The $a$-gauge conditions for all ghost number sectors
are explicitly specified in \cite{Asano:2006hk, Asano:2008iu}.
} is defined by
\begin{eqnarray}
\label{eq:a-gaugePhi1}
 (b_0M+ab_0c_0\tilde Q)\Phi_1=0.
\end{eqnarray}
Here, $a$ is a real parameter and 
the operators $M$ and $\tilde Q$ are specified by an expansion
of the Kato-Ogawa BRST operator $Q_{\rm B}$ with respect to ghost zero
mode: $Q_{\rm B}=\tilde Q+c_0L_0+b_0M$. 
In the case $a=0$, the condition (\ref{eq:a-gaugePhi1}) is
 equivalent to the conventional Feynman-Siegel gauge.
Actually, by investigating the massless sector explicitly in the
 quadratic level of the string field action including spacetime ghost
 fields, the parameter $a$ corresponds to the gauge parameter  $\alpha$
 in the covariant gauge in the ordinary gauge theory as
 $\alpha=1/(a-1)^2$ \cite{Asano:2006hk}. 
In the case $a=\infty$, the condition (\ref{eq:a-gaugePhi1})
is given by $b_0c_0\tilde Q\Phi_1=0$ and corresponds to the Landau
 gauge.

We should note that, in the case $a=1$,
 the condition (\ref{eq:a-gaugePhi1}) is ill-defined
at the free level because it becomes $b_0c_0Q_{\rm B}\Phi_1=0$, which
 can not fix the gauge perturbatively.

\section{Construction of numerical solutions
\label{sec:method}}

Here, we explain our strategy to construct numerical solutions in the
$a$-gauge.
We use an iterative procedure, which was used in the case of the Siegel
gauge in \cite{Gaiotto:2002wy}.
Firstly, as an initial configuration $\Psi_{(0)}$, we take
\begin{eqnarray}
\label{eq:Psi0}
 \Psi_{(0)}&=&\frac{64}{81\sqrt{3}}c_1|0\rangle,
\end{eqnarray}
which is a unique nontrivial solution in the lowest level truncation
in the $a$-gauge.
Then, if we have $\Psi_{(n)}$, we specify the next configuration
$\Psi_{(n+1)}$ by solving following linear equations:
\begin{eqnarray}
&&(b_0M+ab_0c_0\tilde Q)\Psi_{(n+1)}=0,\\
&&{\cal P}(Q_{\Psi_{(n)}}\Psi_{(n+1)}-\Psi_{(n)}*\Psi_{(n)})=0,
\end{eqnarray}
where
\begin{eqnarray}
 Q_{\Psi_{(n)}}\Phi\equiv Q_{\rm
  B}\Phi+\Psi_{(n)}*\Phi-(-1)^{|\Phi|}\Phi*\Psi_{(n)}.
\end{eqnarray}
The first equation is the $a$-gauge condition for $\Psi_{(n+1)}$
and the second one comes from the equation of motion:
\begin{eqnarray}
\label{eq:EOM}
 Q_{\rm B}\Psi+\Psi*\Psi=0.
\end{eqnarray}
${\cal P}$ is an appropriate projection operator
to solve the equations.
In our numerical computation, we take ${\cal P}=c_0b_0$ for
simplicity.
If the above iteration converges to a configuration $\Psi_{(\infty)}$,
it satisfies the $a$-gauge condition and
\begin{eqnarray}
 {\cal P}(Q_{\rm B}\Psi_{(\infty)}+\Psi_{(\infty)}*\Psi_{(\infty)})=0,
\end{eqnarray}
which is a projected part of the equation of motion.
In order to confirm the whole equation of motion (\ref{eq:EOM})
for the converged configuration, we should check the remaining
part:\footnote{
In \cite{Hata:2000bj}, 
this condition in the Siegel gauge is called the BRST invariance 
and investigated for the numerical solution.
}
\begin{eqnarray}
 (1-{\cal P})(Q_{\rm
  B}\Psi_{(\infty)}+\Psi_{(\infty)}*\Psi_{(\infty)})=0,
\label{eq:BRSTinv}
\end{eqnarray}
where $1-{\cal P}=b_0c_0$ in our case.

Actually, we performed the above procedure numerically with 
the conventional level truncation.
We constructed the $a$-gauge numerical solution for various $a$
with $(L,2L)$ and $(L,3L)$-truncation,
where $L$ denotes the maximum level (eigenvalue of $L_0+1$) of the
truncated string field and $2L$ or $3L$ indicates the maximal total
level of the truncated 3-string interaction terms.
Starting from (\ref{eq:Psi0}), we continue the above iterations until
the relative error reaches
$\|\Psi_{(M)}-\Psi_{(M-1)}\|/\|\Psi_{(M)}\|<10^{-8}$,
where $\|(\cdots)\|$ denotes the Euclidean norm with respect to an 
orthonormalized basis.
Then, we find that
$\|{\cal P}(Q_{\rm
  B}\Psi_{(M)}+\Psi_{(M)}*\Psi_{(M)})
\|/\|\Psi_{(M)}\|<10^{-8}$
holds for the obtained configuration.
For various $a$, except for the dangerous region  $a\sim 1$, which is
near to the ill-defined gauge condition perturbatively as we noted in
\S\ref{sec:AKa-gauge},
we find that the configuration reaches this accuracy limit
after ten iteration steps or less.

For each obtained converged configuration, 
we computed the left hand side of (\ref{eq:BRSTinv})
and checked that various coefficients approach zero 
and 
\begin{eqnarray}
 \|(1-{\cal P})(Q_{\rm
  B}\Psi_{(M)}+\Psi_{(M)}*\Psi_{(M)})
\|/\|\Psi_{(M)}\|
\end{eqnarray}
is also vanishing with increasing the truncation level.
Therefore, we have regarded our obtained configurations
as numerical solutions in the $a$-gauge
with respect to the whole equation of motion (\ref{eq:EOM})
and evaluated gauge invariants, action and gauge invariant overlap,
 for them.

\section{Evaluation of gauge invariants
\label{sec:results}}

\subsection{Gauge invariants for the numerical solution in the Siegel
  gauge}

In the case of the numerical solution in the Siegel gauge $b_0\Psi=0$,
which is the case $a=0$ in terms of the $a$-gauge,
computation is easier than the case of other value of $a$.
We performed the numerical computations up to level $L=20$.\footnote{
Calculations for higher truncation levels ($L\geq 18$) were performed by
our C++ and Fortran program.
}

\begin{table}[htbp]
\begin{center}
\begin{tabular}[tb]{|c|c|c|}
\hline 
$L$ &
$2\pi^2g^2S(\Psi)|_{(L,2L)}/V_{26}$
&
$2\pi^2g^2S(\Psi)|_{(L,3L)}/V_{26}$
\\
\hline
\hline
2&0.948553&0.959377\\
\hline
4&0.986403&0.987822\\
\hline
6&0.994773&0.995177\\
\hline
8&0.997780&0.997930\\
\hline
10&0.999116&0.999182\\
\hline
12&0.999791&0.999822\\
\hline
14&1.000158&1.000174\\
\hline
16&1.000368&1.000375\\
\hline
18&1.000490&1.000494\\
\hline
20&1.000562&1.000563\\
\hline
\end{tabular}
\end{center}
\caption{
The value of the action for the numerical solution with $(L,2L)$ and
 $(L,3L)$ truncation in the Siegel gauge.
The values are normalized by the analytic result for Schnabl's
 solution $S(\Psi_{\rm Sch})/V_{26}=1/(2\pi^2g^2)$, which is equal to
 the D-brane tension.
Up to the level $L=18$, the above data are consistent with those in
 \cite{Gaiotto:2002wy}.
\label{tab:action_Siegel}
}
\end{table}
\begin{table}[htbp]
\begin{center}
\begin{tabular}[tb]{|c|c|c|}
\hline 
$L$ &
$2\pi{\cal O}_V(\Psi_{(L,2L)})/V_{26}$
&
$2\pi{\cal O}_V(\Psi_{(L,3L)})/V_{26}$
\\
\hline
\hline
2&0.878324&0.889862\\
\hline
4&0.929479&0.931952\\
\hline
6&0.950175&0.951079\\
\hline
8&0.960617&0.961175\\
\hline
10&0.967790&0.968115\\
\hline
12&0.972321&0.972560\\
\hline
14&0.976005&0.976171\\
\hline
16&0.978544&0.978677\\
\hline
18&0.980802&0.980904\\
\hline
20&0.982432&0.982517\\
\hline
\end{tabular}
\end{center}
\caption{
The value of the 
gauge invariant overlap for the numerical solution with $(L,2L)$ and
 $(L,3L)$ truncation in the Siegel gauge.
The values are normalized by the analytic result for Schnabl's
 solution (\ref{eq:O_eta=1/2pi}).
\label{tab:overlap_Siegel}
}
\end{table}

In Tables~\ref{tab:action_Siegel} and \ref{tab:overlap_Siegel}, 
we show our numerical results.
The values of the action overshoot 100\% of the D-brane tension
for $L\ge 14$ as in  Table \ref{tab:action_Siegel}.
This phenomenon has been reported and expected that
the value will come back to one for further higher level in
\cite{Gaiotto:2002wy}.
On the other hand, the values of the gauge invariant overlap
monotonically approach the analytic value of the Schnabl's solution
as in Table \ref{tab:overlap_Siegel} although the approaching speed is
rather slow compared to the behavior of the action.

Anyway, our results in
Tables \ref{tab:action_Siegel} and \ref{tab:overlap_Siegel}
seem to imply that these (normalized) gauge invariants become
 the value of one for $L\to \infty$.
If so, these give evidence of the gauge equivalence between the numerical
solution in the Siegel gauge and Schnabl's analytic one
\cite{Kawano:2008ry}.

\subsection{Gauge invariants for the numerical solutions in the 
$a$-gauge}

\begin{figure}[htbp]
 \begin{center}
\includegraphics[width=340pt]{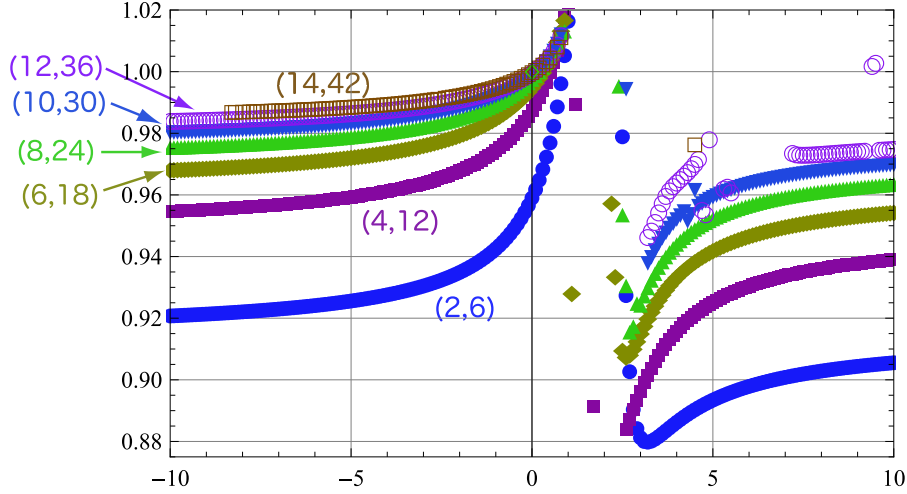}
\end{center}
\caption{Plots of the action
for various $a$-gauge solutions $\Psi_{a,L}$
in the $(L,3L)$ truncation. The
 horizontal axis denotes the value of $a$
and the vertical one denotes 
the normalized action
$2\pi^2g^2S(\Psi_{a,L})/V_{26}$.
The label $(L,3L)$ for each ``curve'' denotes the truncation
level.
}
\label{fig:actionL3L}
\end{figure}

\begin{figure}[htbp]
 \begin{center}
\includegraphics[width=340pt]{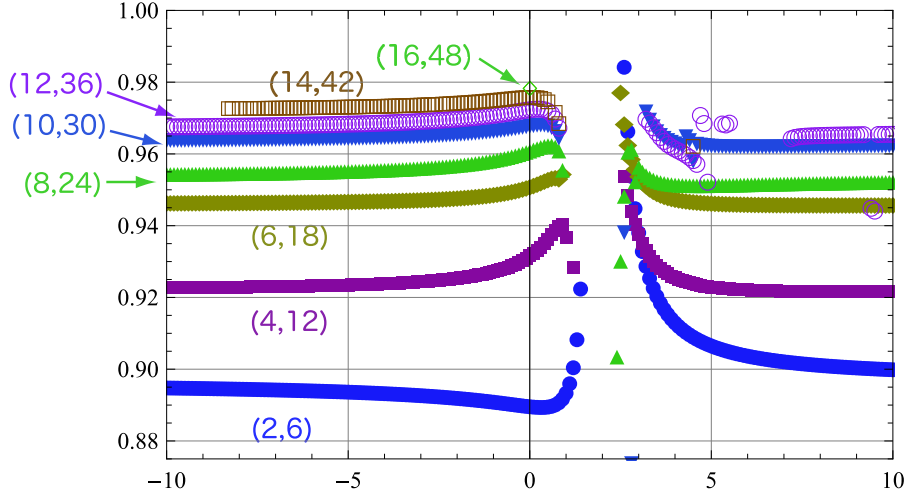}
\end{center}
\caption{Plots of the gauge invariant overlap
for various $a$-gauge solutions $\Psi_{a,L}$
in the $(L,3L)$ truncation. The horizontal axis denotes 
the value of $a$ and the vertical one denotes
the normalized gauge invariant overlap $2\pi{\cal
 O}_V(\Psi_{a,L})/V_{26}$.
}
\label{fig:overlapL3L}
\end{figure}

\begin{figure}[htbp]
 \begin{center}
\includegraphics[width=340pt]{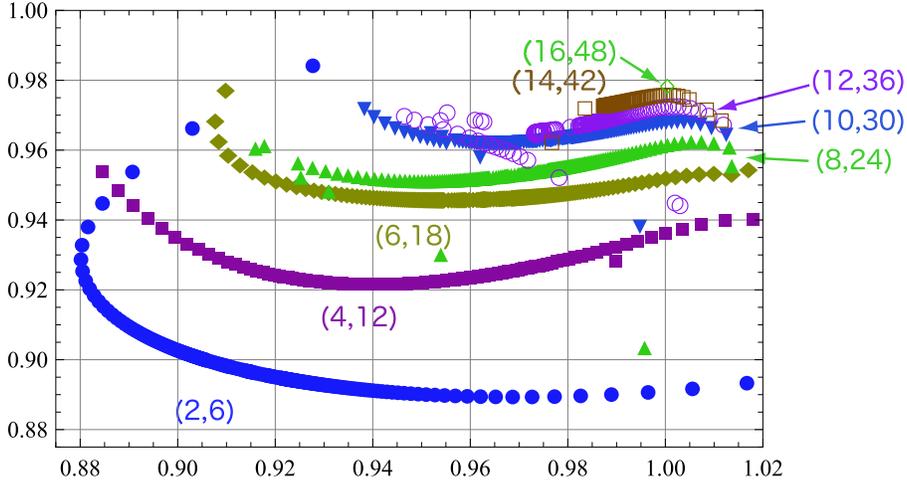}
\end{center}
\caption{Plots of gauge invariants 
for various $a$-gauge solutions $\Psi_{a,L}$
in the $(L,3L)$ truncation. The
 horizontal axis denotes the normalized action
and the vertical one denotes 
the normalized gauge invariant overlap.
Each point denotes the value of $(2\pi^2g^2S(\Psi_{a,L})/V_{26},
2\pi{\cal O}_V(\Psi_{a,L})/V_{26})$
for various $a$ values including $a=\infty$.
 The left part of the ``curve'' for each level
corresponds to $4\lesssim  a < +\infty$ and the right part corresponds to
$-\infty < a \lesssim 1/2$. The plots for $a\to +\infty$ and $a\to
 -\infty$ are continuously connected at that of the Landau gauge
 ($a=\infty$).
}
\label{fig:ginvsL3L}
\end{figure}

Here we show the evaluation of the gauge invariants for numerical
solutions in the $a$-gauge.
Figs.~\ref{fig:actionL3L}, \ref{fig:overlapL3L} and 
\ref{fig:ginvsL3L} are plots for the $(L,3L)$ truncation.\footnote{
Only one datum for $(16,48)$ truncation, which is in the
Siegel gauge $(a=0)$, has been computed.
For other $a$-gauges ($a\ne 0$), calculations are harder in our
{\sl Mathematica} program.
}
Similar tendency of plots is found in the level $(L,2L)$
truncation \cite{Kishimoto:2009cz}.


For various $a$ in the region $a\lesssim 0$, $a\gg 1$, the normalized
gauge invariants, action (Fig.~\ref{fig:actionL3L})
 and gauge invariant overlap (Fig.~\ref{fig:overlapL3L}),
 approach the value of one with increasing level.
The speed of approach to one for the gauge invariant overlap
is slower than that of the action as in the case of the Siegel gauge
(Tables~\ref{tab:action_Siegel}, \ref{tab:overlap_Siegel}).
Although only $a=1$ gauge is ill-defined at the free level, 
interactions are included in the numerical calculations and 
hence the values in the region near $a\sim1$ are unstable. 
In fact, the iterations do not converge in the dangerous
region near $a\sim 1$.

Fig.~\ref{fig:ginvsL3L} shows that both (normalized) gauge invariants
for numerical solutions in the various $a$-gauges tend to converge to
one with increasing truncation level. Namely,
 in the limit $L\to \infty$,
\begin{eqnarray}
&&S(\Psi_{a,L})\to  S(\Psi_{\rm Sch}),~~~~~~
{\cal O}_V(\Psi_{a,L})\to  {\cal O}_V(\Psi_{\rm Sch}),
\end{eqnarray}
are suggested for various $a$ ($-\infty \leqq a \lesssim 0$,
$1\ll a\leqq\infty$).
This seems to imply that not only the Siegel gauge ($a=0$) solution but
also various $a$-gauge solutions constructed as in \S~\ref{sec:method}
 are all gauge equivalent to the Schnabl's analytic solution.

\section{Conclusion
\label{sec:summary}}

We have evaluated gauge invariants (action and
gauge invariant overlap) for numerical solutions in the $a$-gauge by
level truncation.
We have used an iterative method to construct these solutions
and have checked consistency of the equation of motion for them. 
Except for the region at approximately $a=1$, where $a$-gauge condition
becomes ill-defined at the free level, our various solutions in the
$a$-gauge reproduce analytic values of Schnabl's solution for tachyon
condensation. The results are consistent with the expectation that
various solutions in the $a$-gauge, including the Siegel gauge solution
($a=0$), are gauge equivalent to Schnabl's solution.
Therefore, they may represent a unique non-perturbative vacuum, where a
D25-brane vanishes.

\paragraph{Acknowledgements}
This work was supported in part by JSPS Grant-in-Aid for Scientific
Research (C) (\#21540269).
The work of I.~K. was supported in part by a Special Postdoctoral 
Researchers Program at RIKEN.
The work of T.~T. was supported in part by Nara Women's University
Intramural Grant for Project Research.
Numerical computations in this work were partly carried out 
on the Computer Facility of the Yukawa Institute for
Theoretical Physics in Kyoto University
and
the RIKEN Integrated Cluster of Clusters (RICC) facility.

\end{document}